\documentstyle[12pt,epsfig]{article}

\newcommand 	\be   	{\begin{equation}}
\newcommand 	\ee   	{\end{equation}}
\newcommand 	\lan  	{\langle}
\newcommand 	\ran  	{\rangle}
\newcommand 	\MF   	{\scriptscriptstyle {\rm MF}}
\newcommand	{\overl}[1]	{\overline{#1}}
\newcommand	\de	{{\rm d}}
\renewcommand     \b      {\beta}

\begin{document}

\title{ Mean Field Dynamical Exponents \\
	in Finite-Dimensional Ising Spin Glass}

\author{G. Parisi, P. Ranieri, F. Ricci-Tersenghi 
and J. J. Ruiz-Lorenzo\\[0.5em]
{\small Dipartimento di Fisica, Universit\`a di Roma {\em La
Sapienza}}\\ 
{\small and INFN, sezione di Roma I}\\
{\small P. A. Moro 2, 00185 Roma (Italy)}\\[0.3em]
{\small   \tt parisi,ranieri@roma1.infn.it ~
        ricci,ruiz@chimera.roma1.infn.it}\\[0.5em]}

\date{January 8, 1997}

\maketitle

\begin{abstract}

We report the value of the dynamical critical exponent $z$  for the six
dimensional Ising spin glass, measured in three different ways: from
the behavior of the energy and the susceptibility with the Monte Carlo
time and by studying the overlap-overlap correlation function as a
function of the space and time. All three results are in a very good
agreement with the Mean Field prediction $z=4$. Finally we have
studied numerically the remanent magnetization in 6 and 8 dimensions
and we have compared it with the behavior observed in the SK
model, that we have computed analytically.

\end{abstract}  

\thispagestyle{empty}
\newpage

\section{\protect\label{S_INT}Introduction}

Nowadays the only way to calculate analytically critical exponents in
non exactly soluble models is the use of field theory techniques. One
can choose to use the $\epsilon$-expansion~\cite{WILSON} or work at
fixed dimension~\cite{GIORGIO}. 

These approaches have had a great success in the study of the pure
Ising  model (i.e. the $\phi^4$ theory).

In the spin glass case the starting theory is a $\phi^3$ theory in the
limit of zero components of the fields (induced by the replica trick).
Another theories that can be described by the $\phi^3$ field
theoretical description are the percolation, the Potts models and the
Lee-Yang singularities. The upper critical dimension of this theory is
6.

In order to check the validity of the predictions of the
$\epsilon$-expansion (where the calculations are easier than work at
finite dimension) we need to have numerical results on the critical
exponents in dimensions near the upper critical one ($d_u=6$),
i.e. to simulate in 5 and 4 dimensions (where $\epsilon=1,2$).

The previous discussion refers to the static of the system and can be
extended in a easy way to the dynamics of the spin glasses. 

Recently two of the authors (G. Parisi and P. Ranieri)\cite{PARA} has
been able to compute the one loop correction to the dynamical critical
exponent $z$ whose Mean Field (MF) value (the base of this
 $\epsilon$-expansion)
is 4. They found 
\be z(\epsilon)=4\left(1-\frac{\epsilon}{12}\right),
\ee 
where $\epsilon=6-d$.

To test this analytical result is mandatory to do numerical
simulations in five dimensions.

This paper is the first step to the calculation of the dynamical
critical exponent in five dimensions. We can found in the
literature results in four \cite{4DIM} 
and three dimensions \cite{OGIELSKI}.

In this paper we use different techniques in the six dimensional case
(where we have as reference the Mean Field value) to extract the value
of $z$. We plain in the future simulate in five dimensions using the
methodology developed and tested in the present paper.

In particular we obtain the $z$ exponent~\cite{RIEGER} using three
different off-equilibrium ways: the decay of the energy, the growth of
the non linear susceptibility and by means to the scaling of the
overlap-overlap correlation function. These three methods provide us
of three estimates of $z$ or ratios of $z$ with the static critical
exponents in a very good agreement with the Mean Field predictions
($\nu_{\MF}=1/2$, $\eta_{\MF}=0$ and $z_{\MF}=4$).

Whereas the $z$ calculations have been done at the critical
temperature of the system we have performed extra numerical simulation
inside the cold phase to monitorize the ``expected" dependence on the
temperature of the $z$ exponent (like has been obtained in $3$ and $4$
dimensions~\cite{3DIM, 4DIM}) and to check the predictions of De
Dominicis {\em et al.} (monitorizing the growth of the susceptibility)
that predict that the propagator restricted to the $q=0$ ergodic
component goes like $p^{-4}$, where $p$ is the momenta of the
propagator. Obviously at the critical point we expect the usual
dependence on the momenta, i.e. $p^{-2}$.

We have also studied an interesting observable, the remanent
magnetization. This magnetization is defined as follows: we put the
system under a large magnetic field, turn off it and follow the
decay of the magnetization of the system. This observable has a great
importance because its measurement is accessible
experimentally~\cite{GRANSVENOLUNCHEN87}. This decay is like
\be
M(t) \sim t^{-\lambda},
\label{E_rem}
\ee
which define the $\lambda$ exponent~\cite{KISASCHRI96}. Following
the results of D. Fisher and H. Sompolinsky~\cite{FISHSOMP} we expect
that $\lambda$ will be equal to the MF value only for $d \ge 8$,
i.e. for exponents which are reminiscent of the ``magnetic field" the
upper critical dimension is eight, not six.

We check this fact numerically and analytically in the present paper
obtaining the value of the $\lambda$ exponent in six, eight and
infinite dimensions (the SK model).

\section{\protect\label{S_NS}Numerical simulation and observables}

We have simulated the $\pm 1$ six dimensional Ising spin glass, whose
Hamiltonian defined in a hypercubic of volume $L^6$ is
\begin{equation}
{\cal H}=-\sum_{<i,j>} S_i J_{ij} S_j,
\end{equation}
where $<i,j>$ denotes sum to nearest neighbor pairs, $J_{ij}=\pm 1$
(with the same probability) are quenched variables and $S_i=\pm 1$ are
spin variables. 

We have simulated $L=8$ and $L=10$ sizes at the critical point with a
number of samples respectively of 103 and 13, and also 106 samples of
an asymmetric system of size $12 \times 8^5$.

The static of this model was studied by Wang and
Young~\cite{WANG_YOUNG}.
Simulating lattice sizes up to $L=8$ they found that the 
static critical exponents were compatibles with the Mean Field
predictions ($\nu_{\MF}=1/2, \eta_{\MF}=0$) and
that there were 
logarithmic corrections to the Mean Field exponents, effect linked to
the upper critical dimension. Their estimate for the critical
temperature was $T_c = 3.035 \pm 0.01$.

The main aim of this paper is to measure the dynamical critical
exponent $z$ in order to compare it with the Mean Field results
($z_{\MF}=4$). To do this we have measured the behavior of the energy
and susceptibility as a function of the Monte Carlo time
\be
E(t)-E_\infty \propto t^{-\delta}
\ee
\be
\chi(t) \propto t^h
\ee
and the $q-q$ correlation function.

The first observable that we will examine is the dependence of the
energy with the Monte Carlo time. We assume that at the critical point
(and only at the critical point) it is possible to connect the
approach to equilibrium of the energy and of the equal time
correlation functions to the equilibrium static and dynamical
exponents. For example in the case of the energy we find:
\be
E(t)-E_\infty \propto t^{-{\rm dim}(E)/z},\;\;\;\; T=T_c, 
\ee 
where $z$ is the dynamical critical exponent, ${\rm dim}(E) \equiv
d-1/\nu$ is the dimension of the energy operator and $d$ is the
dimensions of the space. Assuming $d=6$ and \mbox{$\nu = \nu_{\MF} =
1/2$} we have that the exponent of the energy decay at $T=T_c$ is
$\delta = 4/z$.

Analogously we can write for the non linear susceptibility
\be
\chi(t)=L^d {\overline{\lan q^2(t) \ran}},\;\;\;\; q(t) =
\frac{1}{L^d} \sum_i \sigma_i(t) \tau_i(t),
\ee
where $\sigma$ and $\tau$ are two real replicas with the same quenched
disorder, the following dependence on the Monte Carlo time
\be
\chi(t) \propto t^{(2-\eta)/z}, \;\;\;\; T=T_c,
\ee
as $t \ll \tau_{\rm eq}(L)$, where $\tau_{\rm eq}(L)$ is the
equilibration time, which should diverge as $\tau_{\rm eq}(L) \propto
L^z$. Where we have used that ${\rm dim}(\chi)=2-\eta$. Assuming the
Mean Field value for $\eta$ we have that the exponent of the
susceptibility is $h = 2/z$ at $T=T_c$. 

From these formul\ae~it is possible to calculate the dynamical
exponent via two different observables. In the six dimensional
case, if $z = z_{\MF} = 4$  we should expect a behavior like $t^{-1}$
for the energy and like $t^{1/2}$ for the non linear susceptibility.

A third way to calculate the dynamical exponent is using the
overlap-overlap correlation length at distance $x$ and time $t$
defined by 
\be
G(x,t)\equiv \frac{1}{V} \sum_i \overl{\langle \sigma_{i+x} \tau_{i+x}
 \sigma_i \tau_i \rangle },
\label{E_q-q_def}
\ee  
where again $\sigma$  and $\tau$  are two  real replicas with the
same disorder.
In the simulations we start from a random configuration ($T=\infty$)
and suddenly we quench the system to $T_c$ or below.
Then the system begins to correlate itself and we can define a time
dependent off-equilibrium correlation length, $\xi(T,t)$, as the
typical distance over which the system have already developed
correlations different from zero, i.e. $G(x,t) \simeq 0$ for
$x > \xi(T,t)$.
The growth of this correlation length with the Monte Carlo time
defines the dynamical exponent $z$ trough
\be
\xi(T,t) \propto  t^{1/z(T)}.
\ee

We have seen that in three and four dimensions \cite{3DIM,4DIM}
the data fit very well
the following functional form
\be
G(t,x) = \frac{A(T)}{x^\alpha} \exp \left\{-\left( \frac{x}{\xi(T,t)}
\right)^\gamma  \right\} ,
\label{E_q-q_cor}
\ee
Thereby, this will be the third way to obtain the dynamical critical
exponent. This third estimate of $z$ is independent of the values of
the static critical exponents.

The equilibrium overlap-overlap
correlation function constraint to $q=0$ was obtained by De Dominicis
{\em et al.}~\cite{DE_DOMINICIS}, which in 6 dimensions reads
\be
C_{\rm SRSB}(x)|_{q=0} \sim 
                \left\{ \begin{array}{ll}
			x^{-4}           & \mbox{ if   $T=T_c$,}\\
			x^{-2}           & \mbox{ if   $T<T_c$.}
		      \end{array}
                 \right.
\label{chi_rsb}
\ee
The fact that the equilibrium correlation function $C(x)$ has a power
law decay also at $T<T_c$, implies that spin glasses are always
critical in the glassy phase (i.e. below $T_c$) and therefore we can
relate the off-equilibrium behavior of the correlation function to the
equilibrium critical exponents. Using that the susceptibility is the
integral of the correlation function
\be
\chi=\int {\rm d}^6 x\; C(x) 
\label{chi_def}
\ee
and that in the region where the susceptibility grows, following a
power law of the time, the overlap is very small,  we can
use equations (\ref{chi_rsb}) in equation (\ref{chi_def}), obtaining
\be
\chi(t) \sim 
                \left\{ \begin{array}{ll}
			t^{1/2}           & \mbox{ if   $T=T_c$,}\\
			t^{4/z(T)}          & \mbox{ if   $T<T_c$.}
		      \end{array}
                 \right.
\ee
If we take the limit $T\to T_c$ in the above equation we obtain that
$h(T)$, the exponent of the growth of the susceptibility, must be
discontinuous at the critical point (i.e. $h(T_c^-)=1$ while
$h(T_c^+)=1/2$).

Moreover, if we assume that $1/z(T)$ is proportional to the
temperature (this happens in 3 and 4 dimensions~\cite{3DIM, 4DIM}) we
must obtain a linear dependence on the temperature for $h(T)$ in the
low temperature phase.

Finally we will study the decay of the remanent magnetization defined
as
\be
M(t,t_w)=\frac{1}{L^d}\sum_{i=1}^{L^d}\sigma_i(t) \sigma_i(t_w),\;\;\;
t \gg t_w.
\label{E_M}
\ee
This observable decays like
\be
M(t,t_w) \propto t^{-\lambda}.
\ee
We find that the Mean Field prediction for this exponent is
$\lambda=5/4$.

\section{\protect\label{S_RES}Results in $d=6$}

The great part of the simulation work have been done at the critical
temperature, chosen as the weighted mean between the one found by Wang
and Young~\cite{WANG_YOUNG} ($T_c = 3.035 \pm 0.01$) and the one
calculated by the series expansion~\cite{SERIES_EXP} ($T_c = 3.027 \pm
0.005$): $\beta_c = 0.3302 \pm 0.0005$.  In particular we have tested
that the exponents we measure do not vary in a significant way if the
temperature is changed by an amount of same order than the uncertainty
on $T_c$.  To this purpose we have simulated the same system (of
volume $8^6$) at the inverse temperatures $\beta_1=0.330$ and
$\beta_2=0.331$, checking that the dynamics were compatible.

Verified that, for the range of time and sizes we have used, the
exponents we are interested in do not depend on the precise $T_c$
choice, we have decided to run all the subsequent simulations at
$\beta_c = 0.330$. At the critical point we have simulated more than
one hundred samples of size $8^6$ and 13 samples of size $10^6$. The
last number of samples may appear too small to average over the
disorder; in fact we have used mainly the data from the $8^6$ systems
to calculate the moments of the distribution of the overlaps. The
data from the bigger systems have been used to measure
almost-self-averaging quantities like energy whose fluctuations are
very small considering that we are working with system with a million
of spins.

\begin{figure}
\begin{center}
\leavevmode
\centering\epsfig{file=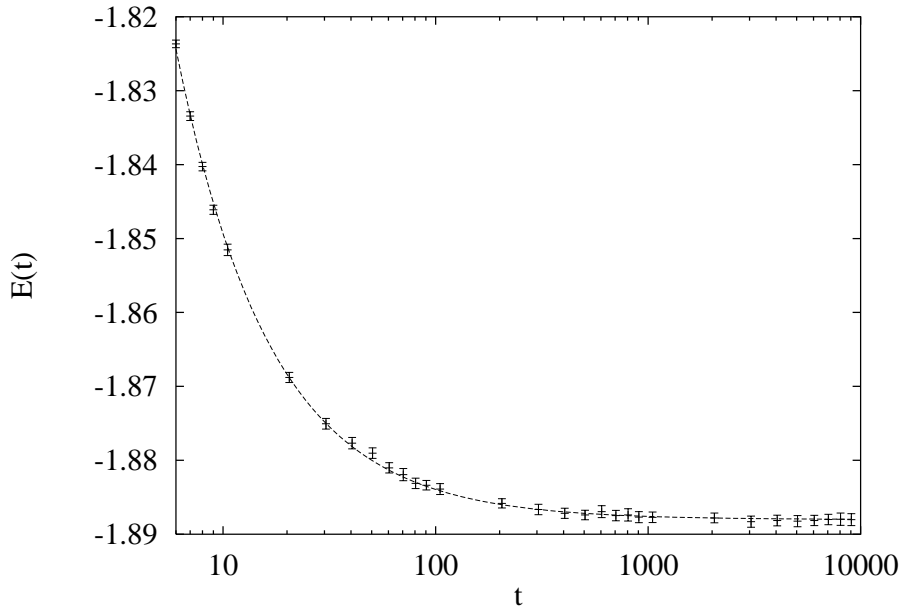,width=0.75\linewidth}
\end{center}
\protect\caption{Energy decay in a $10^6$ system at $T=T_c$.
\protect\label{F_ener}}
\end{figure}

\begin{figure}
\begin{center}
\leavevmode
\centering\epsfig{file=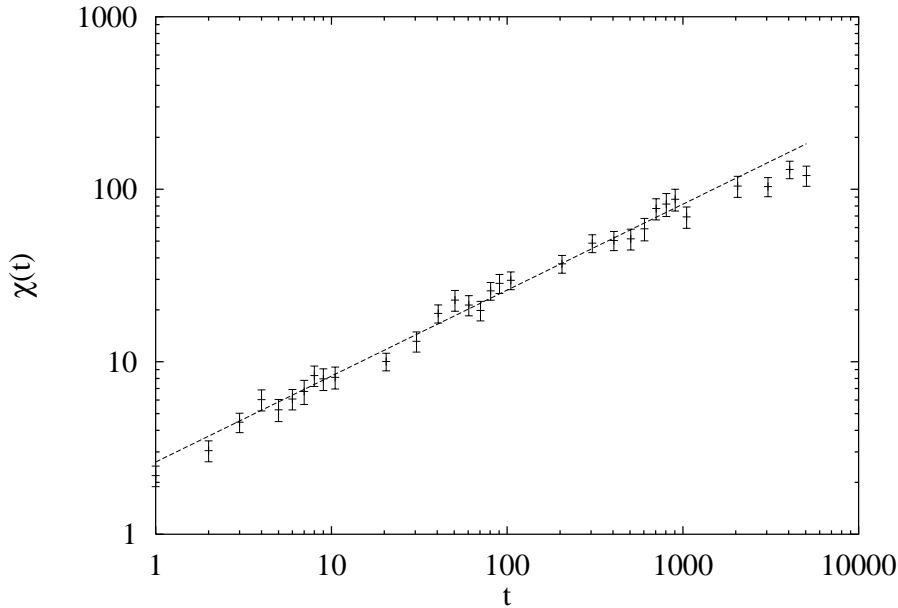,width=0.75\linewidth}
\end{center}
\protect\caption{Susceptibility growth in in systems of size $8^6$ at
$T=T_c$. \protect\label{F_susc}}
\end{figure}

The results obtained are shown in figure~\ref{F_ener} for the energy
decay and in figure~\ref{F_susc} for the non linear susceptibility
growth.
 
We have tried to fit the energy decay both with a power law ($E(t) =
E_\infty + A t^{-\delta}$) and a logarithmic law ($E(t) = E_\infty + A
[\ln(t/\tau)] ^ {-\delta}$).  We are interested in the asymptotic
behavior of the decay; then we fit the data in the range $t \in
[t_{min}, \infty)$ for various choices of $t_{min}$ and we expect that
the parameters of the fit converge rapidly when we increase $t_{min}$.
The impossibility of fitting all the data with a single law (for
$t<6$) is due to the existence of an initial short time regime of a
few steps during which the dynamic does not follow yet the asymptotic
behavior.  We find that the logarithmic law do not describe well the
data because, even if it has more adjustable parameters, the best
values of the parameters depend strongly on $t_{min}$, they are very
correlated and they tend to unphysical values.  On the other hand we
find that fitting with the power law the values of the parameters
$E_\infty$, $A$ and $\delta$ converge to a stable value, with
$t_{min}$ of order of few MCS.  In figure~\ref{F_ener} we plot the
power fit obtained with $t_{min}=6$ (this is the lowest value for
which the fit holds the $\chi^2$ test); the best parameters are:
$E_\infty = -1.8880 \pm 0.0001$, $A = 0.37 \pm 0.01$ and $\delta =
-0.98 \pm 0.01$.  We note that the decay exponent is compatible with
the mean field value ($\delta_{\MF} = -1$).
 
The line plotted in figure~\ref{F_susc} is the best power fit to the
susceptibility data.  We have to take care when we try to interpolate
in such a way because we know that the susceptibility growth follows a
power law only in a limited time window; in fact at the beginning of
the simulation the dynamics needs some short time to reach the
asymptotic regime~\footnote{This initial time increases when the
temperature is lowered.} and then because of the finite size of the
system they have to converge to some finite value, i.e. the data of
figure~\ref{F_susc} have to converge to a plateau.  These two effects
may induce systematic deviations in the estimation of the $h$
exponent: $\chi(t) \propto t^h$.  In our case the first transient is
almost absent thanks to the sufficiently high temperature, but
problems may arise from the existence of a plateau.  In fact we find
that fitting all the data points with the power law the $\chi^2$ value
is not acceptable while, just discarding the last few points, the fit
works very well.  The line reported in figure~\ref{F_susc} is the
power law obtained fitting the data in the range $t \in [6, 2000]$
\footnote{The fits done in different time windows give similar
results.}; we note that the last three points are below the best fit,
which is probably due to the effect just described.  Our estimation of
the dynamical exponent $h$ is $h = 0.49 \pm 0.02$ which is compatible
with the mean field value ($h_{\MF} = 1/2$).

\begin{figure}
\begin{center}
\leavevmode
\centering\epsfig{file=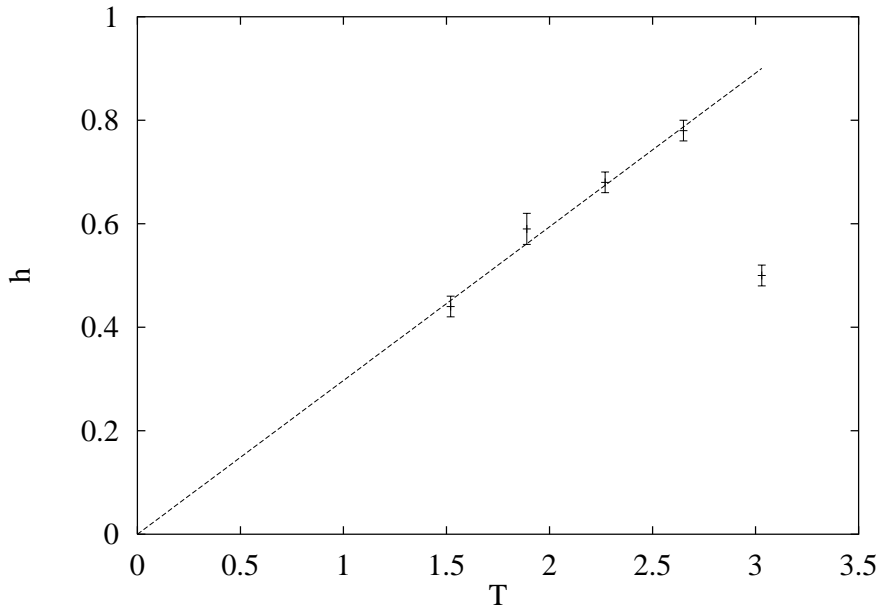,width=0.75\linewidth}
\end{center}
\protect\caption{$h(T)$ exponent vs. $T$ in the region $T \le
T_c$. \protect\label{F_h_T}}
\end{figure}

We also show in figure \ref{F_h_T} the results (all obtained on a $8^5
\times 12$ lattice and with 200 samples) for the $h$ exponent in the
low temperature phase ($T/T_c = 0.5, 0.625, 0.75, 0.875$).  We also
plot in this figure the value that we have obtain at $T_c$
($h=0.49$). It is clear that $h(T)$ is a discontinuous function at the
critical point and that the limit from below, assuming a linear
behaviour, ($h(T_c^-) \simeq 0.9$) is almost twice the value of
$h(T_c^+)=0.5$, in very good agreement with the correlation functions
found by De Dominicis {\em et al.}~\cite{DE_DOMINICIS}. The quite
small discrepancy can be due to the crossover between the two regimes
in a finite lattice or to logarithmic corrections.

Moreover we can see that the dependence of this exponent is well
described by a linear law of the temperature, according to a $z(T)$
inversely proportional to $T$.
 
\begin{figure}
\begin{center}
\leavevmode
\centering\epsfig{file=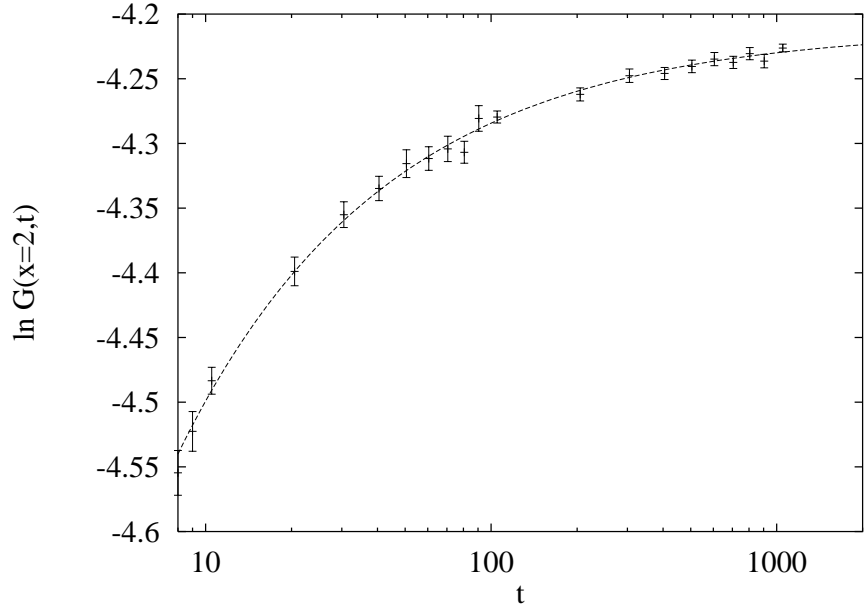,width=0.75\linewidth}
\end{center}
\protect\caption{$\ln G(x\!=\!2,t)$ vs. $t$ for the asymmetric
system at $T=T_c$. \protect\label{F_G_t}}
\end{figure}

Measured the equal time $q-q$ correlation functions defined in
eq.(\ref{E_q-q_def}), we have studied the spatial correlations by the
following technique (already used with success for the data analysis
in 3 and 4 dimensions~\cite{3DIM, 4DIM}). We expect a functional
formula for this correlation function of the form
\be 
G(t,x) = \frac{a(T)}{x^\alpha} \exp \left\{-\left( \frac{x}{\xi(T,t)}
\right)^\gamma \right\},
\label{E_G_tot}
\ee
where, as usual, $\xi(T,t) \propto t^{1/z(T)}$ is the dynamical
correlation length. For each value of the distance we fit the data of
the $q-q$ correlation function by the formula 
\be
G(x,t) = G_\infty(x) \exp \left[ A(x) t^{-B} \right] \;\;\;\;\;\;
\forall \;\; {\rm fixed} \;\; x,
\label{E_G_t}
\ee
verifying that the value of the $B$ parameter is almost independent
from $x$ and then fixing it during the following study.
The dynamical exponent can be expressed as the ratio $z=\gamma/B$,
where $\gamma$ may be estimated by the power law fit $A(x) \propto
x^\gamma$. By this way also our third estimation $z=4.2 \pm 0.2$ is
compatible with the Mean Field value.
In figure~\ref{F_G_t} we have plotted  $\ln G(x\!=\!2,t)$
versus $t$, together with the best fit (eq.(\ref{E_G_t})).

In the infinite time limit the function $G(x,t)$ converges to
$G_\infty(x)$ which must give information on the $q-q$ correlation
function at zero overlap, calculated for the SK model by De Dominicis
{\em et al.}~\cite{DE_DOMINICIS}.
They found that $G_{q=0}(x) \propto x^{-4}$ at the critical
temperature.

\begin{figure}
\begin{center}
\leavevmode
\centering\epsfig{file=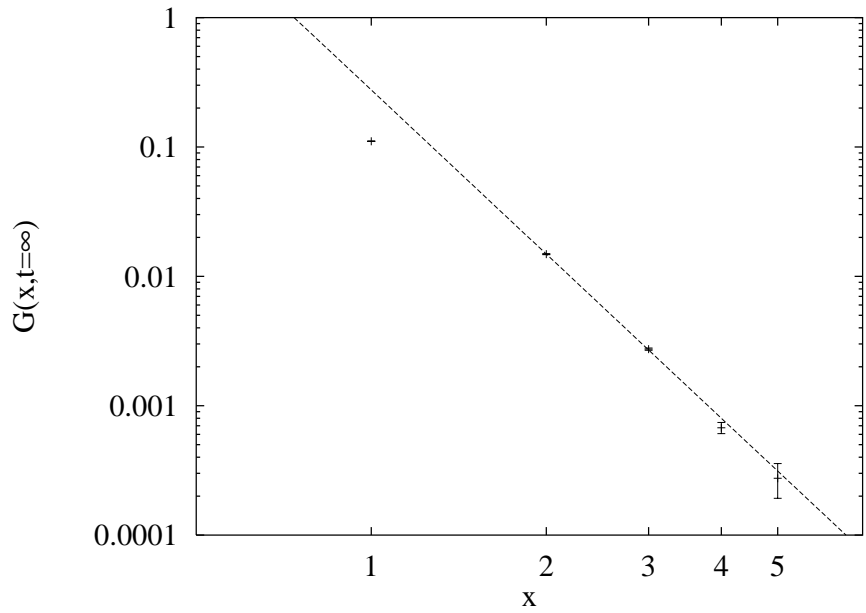,width=0.75\linewidth}
\end{center}
\protect\caption{$G_\infty(x)$ for the partially asymmetric system at
$T=T_c$. \protect\label{F_G_x}}
\end{figure}

By our simulation we find the $G_\infty(x)$ plotted in
figure~\ref{F_G_x} together with the best power fit in the range $x
\in [2,5]$.
We note that the point in $x=1$ if far away from the fit because for
$\ln(x) \rightarrow -\infty$ the data must converge to 1 (by
definition $G(x\!=\!0)=1$) while the power fit diverges.
Anyway we are interested in the asymptotic behavior which seems to be
well described by a power law, $G_\infty(x) \propto x^{-\alpha}$ with
an exponent $\alpha = 4.2 \pm 0.1$, in agreement with the Mean Field
result previously cited.

\section{\protect\label{S_MAG}Remanent magnetization}

The last part of this study has been dedicated to the decay of the
remanent magnetization.  We prepare the system with all the spins up
($M(t\!=\!0) = 1$) and then we let it evolve toward the equilibrium
where, in absence of any external field, no magnetization have to
remain. In the cold phase we expect the decay of the remanent
magnetization to be of algebraic type (see equation (\ref{E_rem}));
in particular we are interested in the exponent of such a decay at the
critical temperature (which hereafter will be called simply
$\lambda$), to compare it with the same exponent of the SK model.

We report in the following subsections our calculation of the
$\lambda$ exponent in the MF approximation (see reference \cite{CUDE}
for another calculation of the $\lambda$ exponent in the spherical
spin-glass model), together with the numerical verification and the
estimations of such an exponent in the finite dimensional cases ($d=6$
and $d=8$) and also for the SK model.

\subsection{\protect\label{S_MAGANA}Analytical and numerical results
in the SK model}

To study analytically the dynamical properties of the SK model we
define the following soft-spin Hamiltonian:
\be
\b{\cal{H}} = -\b \sum_{ij} J_{ij} s_i s_j + \frac{1}{2} \sum_i s_i^2
+ \frac{1}{4!} g \sum_i s_i^4,
\label{HAM}
\ee
where $s_i$ ($i=1,...,N$), are one-dimensional real variables and
$J_{ij}$ is a symmetric matrix with independent elements following a
Gaussian distribution with zero mean and variance proportional to
$1/N$. 
From the random matrix theory \cite{Me} we know that, in the
thermodynamic limit (i.e. $N$ goes to infinity), the probability
distribution for the eigenvalues of $J_{ij}$ is given by the
semi-circle law:
\be
\sigma(\mu) = \frac{1}{2\pi} (4-\mu^2)^{1/2} \ \ \ \ \ \ |\mu|<2.
\ee 
A relaxation dynamics is introduced by the Langevin equation:
\be
\partial_t s_i(t)= - \frac{\partial(\b{\cal{H}})}{\partial s_i(t)} +
\xi_i(t).
\ee
where $\xi_i(t)$ is a Gaussian noise with zero mean and variance 
$\lan\xi_i(t)\xi_j(t')\ran=2\delta_{ij}\delta (t-t')$.
To study the dynamical evolution of this model, we diagonalize the
$J_{ij}$ matrix and we consider the dynamics of the projections
$s^n(t)$ of the spin $s_i(t)$ on the eigenvector directions $\psi_i^n$
(with eigenvalues $\mu_n$), such as $s_i(t)=\sum_{n}
s^{n}(t)\psi_i^{n}$, where $n=1,..,N$ is the eigenvector index.
The properties of independence and orthonormality of the eigenvectors,
\cite{Me} let us define the following Langevin equation for the
component $s^n$:
\be
\partial_t s^n(t) = (\b\mu_n-1) s^n(t) - \frac{g}{3!}
\sum_{\alpha\beta\gamma} s^{\alpha}(t) s^{\beta}(t) s^{\gamma}(t)
\sum_i \psi_i^{\alpha} \psi_i^{\beta} \psi_i^{\gamma} \cdot \psi_i^{n}
+ \xi^n(t).
\ee
In the thermodynamic limit, as shown in \cite{Ra}, the Hartree-Fock 
approximation is exact and we can linearize the Langevin equation:
\be
\partial_t s^n(t) = (\b\mu_n-1) s^n(t) - \frac{g}{2} C(t,t) s^{n}(t) +
\xi^n(t) 
\label{s}
\ee
where $C(t,t)$ is the dynamical auto-correlation,
\be
C(t,t')=\overl{ \lan s_i(t) s_i(t') \ran },
\ee 
evaluated for $t=t'$.
We represent the average over the thermal noise as 
$\lan (\cdot \cdot \cdot) \ran$,
while $\overl{(\cdot \cdot \cdot)}$ indicates the average over the 
disorder as usual. 

With respect to the eigenvalues the auto-correlation function can be
defined as follows
\be
C(t,t')=\int \de\mu\,\sigma(\mu) \lan s^{\mu}(t) s^{\mu}(t') \ran.
\ee
The self-consistent solution of the eq. (\ref{s}) is
\be
s^n(t) = s^n(0) e^{(\b\mu_n-1)t} e^{-g/2\int_0^t \de t'\,C(t',t')} +
\int_0^t \de t''\,e^{(\b\mu_n-1)(t-t'')} e^{-g/2\int_{t''}^t
\de t'\,C(t',t')}\xi^n(t''),
\label{s1}
\ee
where $t=0$ is the initial time.

We want to analyze the evolution of the system at $T=T_c$
from a uniform initial condition: $s^{n}(0)=1 \; , \; \forall n$.
From (\ref{s1}) we obtain the following self-consistent equation for
$C(t,t)$:
\be
C(t,t) = H^2(t) \int \de\mu\,\sigma(\mu) e^{2(\b\mu-1)t} + 2
\int_0^t \de t'' \left( \frac{H^2(t)}{H^2(t'')} \right) \int
\de\mu\,\sigma(\mu) e^{2(\b\mu-1)(t-t'')}
\label{c}
\ee
where
\be
H(t) = e^{-g/2 \int_0^t \de t'\,C(t',t')}.
\ee
Let us suppose for $H(t)$, at $T=T_c$, a time-dependent
asymptotic behavior ($t \to \infty$) like:
\be 
H(t) \sim t^\rho e^{-gt},
\label{g}
\ee
where $\rho $ can be determined  auto-consistently from
eq.(\ref{c}). This hypothesis implies for $C(t,t)$, at large $t$, the
following behavior
\be
C(t,t) \sim \left(2+\Delta T_c\right)- \frac{a}{t}.
\label{c1}
\ee
where $a$ is an opportune constant.  The critical temperature of the
Hamiltonian (\ref{HAM}) is given by $T_c=T_c^0+\Delta T_c=2+\Delta
T_c$ ($T_c^0=2$ is the critical temperature of (\ref{HAM}) when $g=0$)
and that a perturbative  calculation gives $\Delta T_c=-2 g$ and so $2
\beta_c=1+g$ \cite{Ra}.
 
We recall that for $T\sim T_c$ the bigger contribution to the
dynamical relaxation of the spins comes from the region of the maximum
eigenvalue, $\mu=2$.

At the critical point, the first term on the right side of the
eq. (\ref{c}) scales with a power law
$t^{-3/2+2\rho }$. To be consistent with the hypothesis (\ref{g}) and
(\ref{c1}) it should be $\rho\leq 1/4$.

\begin{figure}
\begin{center}
\leavevmode
\centering\epsfig{file=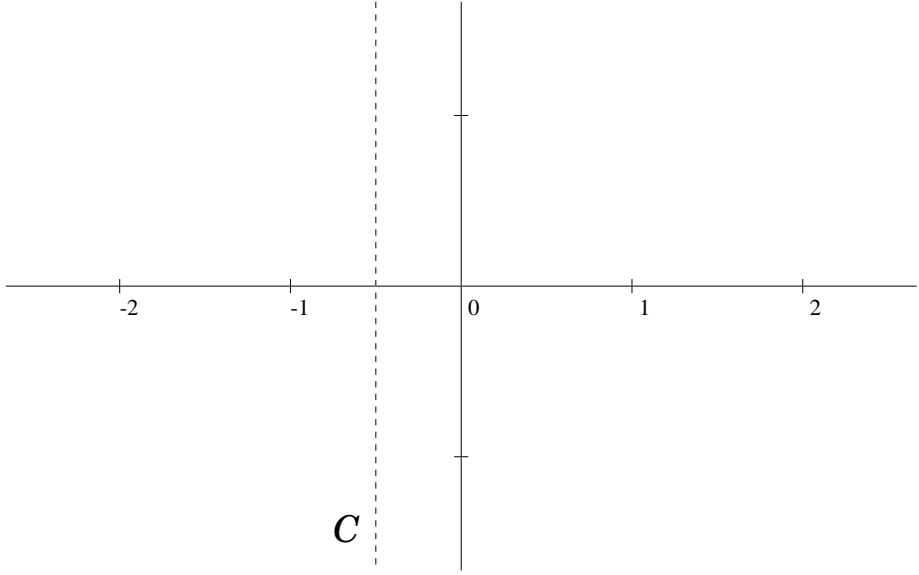,width=0.75\linewidth}
\end{center}
\protect\caption{Integration path. \protect\label{F_path}}
\end{figure}

Now, we have to estimate the second term, which is proportional to:
\be
2\int_0^t \de t''\int \de\mu\,\sigma(\mu) 
\frac{t^{2\rho}}{t''^{2\rho }} e^{-2\b_c(2-\mu)(t-t'')}
\ee
If we consider $t'=t-t''$ and we define the exponential as integral in
the complex plane,
\be
e^{-2\b_c(2-\mu)t'} = \sum_k \frac{(-1)^k}{k!} \left[2\b_c(2-\mu)
t'\right]^k = \int_{{\cal{C}}} \frac{\de s}{2 \pi i} 
\,\Gamma(-s) \left[2\b_c(2-\mu)t'\right]^s,
\ee
where ${\cal{C}}$ is the path shown in figure~\ref{F_path}
(i.e. $s=s_0+i r$, where $r \in {\cal R}$, and $s_0$ whatever real
number in $(-1,0)$) and $\Gamma(s)$ is the Euler Gamma function, we
left with:
\be 
2\int_{{\cal{C}}} \frac{\de s}{2 \pi i}
\, \int_0^t \de t'\,\Gamma(-s)
\frac{t^{2\rho}}{(t-t')^{2\rho}} t'^s \int \de\mu\,\sigma(\mu)
(2\b_c)^s(2-\mu)^s ,
\ee
 
After the integration over $t'$ and $\mu$ we obtain: 
\be
\frac{8}{\sqrt{\pi}}\int_{{\cal{C}}} \frac{\de s}{2 \pi i}
\,\Gamma(-s) t^{1+s}
\frac{\Gamma(s+1)\Gamma(1-2\rho)}{\Gamma(s-2\rho+2)}
4^s(2\b_c)^s\frac{\Gamma(s+3/2)}{\Gamma(3+s)}
\label{r}
\ee
To evaluate the integral (\ref{r}) we analytically continue the
function on the left of the path ${\cal{C}}$, i.e. in the region
where ${\rm Re}\; s<s_0$. Thereby, 
we have to consider the residues of the poles in this region.
The residue of the pole  in $s=-1$ gives the constant
contribution to the auto-correlation function $C(t,t)$ while the
time-dependent behavior, for large $t$, is determined by the value of
$\rho$. In fact, for $\rho=1/4$,  $\Gamma(s+3/2)$ is simplified by
$\Gamma(s-2\rho+2)$, and, for large $t$, the
leading behavior of $C(t,t)$ comes from the residue of the pole
at $s=-2$:
\be
C(t,t) \sim {\rm const} + t^{-1},
\ee
consistently with the hypothesis (\ref{g}) and (\ref{c1}).

For $\rho\neq 1/4$, on the contrary, we do not manage to cancel the
pole at $s=-3/2$ and we should obtain 
\be
C(t,t)\sim {\rm const} + t^{-1/2},
\ee
in contrast with the previous hypothesis. 
Thus the only consistent solution for $H(t)\sim t^\rho e^{-t}$ is
$\rho=1/4$.   

An other way we can obtain the value of the $\rho$ exponent is by
solving the eq. (\ref{c}) in the Laplace transform.
In terms of the function $g(t)$
\be
g(t)\equiv e^{-2gt}\frac{C(t,t)}{\Gamma^2(t)},
\label{tilde}
\ee
the eq. (\ref{c}) at $T=T_c$ becomes:
\be
g(t)=\int \de\mu\,\sigma(\mu) e^{-2\b_c(2-\mu)t} + 2 \int_0^t 
\de\, t''\,
\frac{g(t'')}{C(t'',t'')}\int \de\mu\, \sigma(\mu) 
e^{-2\b_c(2-\mu)(t-t'')}
\ee
By using the asymptotic form (\ref{c1}) of $C(t,t)$, we obtain the
following asymptotic equation for the Laplace transform of $g(t)$,
that we will denote $\tilde{g}(s)$:

\begin{eqnarray}
\tilde{g}(s) & = & \int \de\mu\,\sigma(\mu)
\left[\frac{1}{s+2\b_c(2-\mu)}\right]\\ 
 & + &
\int \de\mu\,\sigma(\mu) \left[\frac{1}{s+2\b_c(2-\mu)}\right] 
\left[\left(1-\frac{\Delta T_c}{2}\right)\tilde{g}(s)+
\frac{a}{2}\int_s^{\infty}\de x\,\tilde{g}(x)
\right]
\end{eqnarray}
By averaging over the eigenvalue distribution we obtain:
\be
\tilde{g}(s) = \left( \frac{1}{2\b_c} - \frac{\sqrt{2s}}{4\b_c^{3/2}}
\right) + \left( \frac{1}{2\b_c} - \frac{\sqrt{2s}}{4\b_c^{3/2}}
\right) \left[ \left( 1-\frac{\Delta T_c}{2} \right) \tilde{g}(s) +
\frac{a}{2} (1-\Delta T_c) \int_s^{\infty} \de x\,\tilde{g}(x) \right]
\label{gamma}
\ee

By remembering that $\beta_c=1/(2+\Delta T_c)$, we can developed the
previous equation in $\Delta T_c$. We will also assume that in the
limit $s \to 0$;  $\int_s^{\infty} \de x\,\tilde{g}(x)$ is negligible 
with respect to $\tilde{g}(s)$, and we finally obtain
\be
\tilde{g}(s)\sim \frac{1}{\sqrt{s}} +{\rm O}(1).
\ee
thus, for $t\rightarrow \infty$, $g(t)\sim 1/t^{1/2}$ and from 
(\ref{tilde}) we obtain $\rho=1/4$. This solution implies that  
$\int_s^{\infty} \de x\,\tilde{g}(x)\ll \tilde{g}(s)$.

At this point, we can determine the decay rate of the correlation
between the system and the initial conditions, i.e. the scaling decay
of the remanent magnetization:
\begin{eqnarray}
M(t)&=& C(t,0) = \overl{\lan s_i(t) \ran}
=\int \de \mu\,\sigma(\mu)\lan s_\mu(t)\ran \\
&=&\int \de\mu\,\sigma(\mu) H(t) e^{(\beta \mu-1)t} 
\sim \frac{1}{t^{3/2}} t^{1/4} \sim t^{-5/4}.
\end{eqnarray}
Then, the analytical prediction for the exponent $\lambda$, such as
$M(t)\sim t^{-\lambda}$, is, in mean field limit, $\lambda=5/4$. 

The result is quite different from the one obtained in the spherical
model at $T=T_c$ where $\lambda=3/4$~\cite{CUDE}.

\begin{figure}
\begin{center}
\leavevmode
\centering\epsfig{file=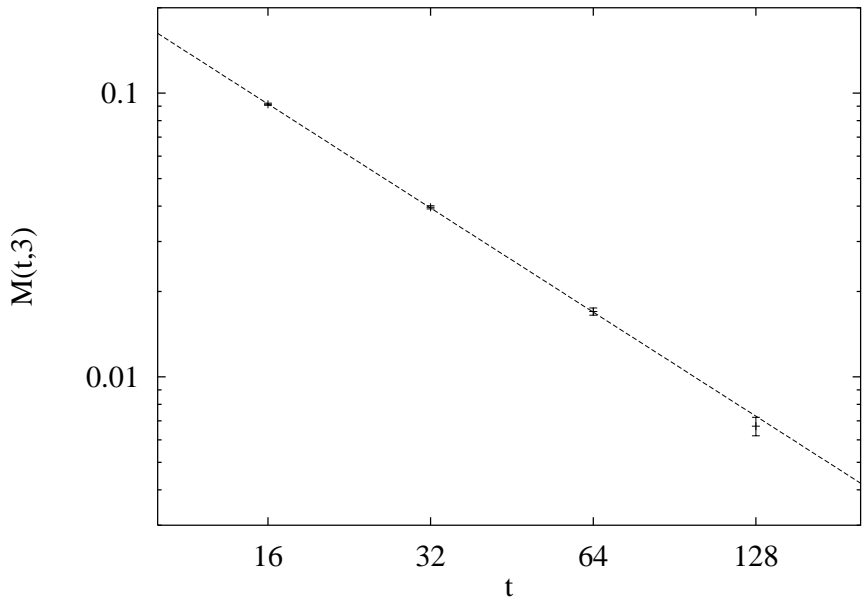,width=0.75\linewidth}
\end{center}
\protect\caption{Remanent magnetization in the SK model at $T=T_c$.
\protect\label{F_mag_sk}}
\end{figure}

For a numerical confirmation of this result we have simulated three SK
models at criticality ($T_c = 1$) of sizes $N=480,992,2016$ with
number of samples of 10000, 5000 
and 1000 respectively, obtaining three estimations of the
$\lambda_{\MF}$ exponent all compatible with the theoretical
prediction.
Since the data for the remanent magnetization have non evident
finite-size effects, we have plotted if figure~\ref{F_mag_sk} the data
averaged over all the simulated samples.
The observable we have measured is $M(t,t_w\!=\!3)$ defined in
eq.(\ref{E_M}), which follows the same decay of $M(t)$ but has some
advantages as will explain below.
The line in figure~\ref{F_mag_sk} is the best power fit which gives
$\lambda_{\MF} = 1.22 \pm 0.02$.

\begin{figure}
\begin{center}
\leavevmode
\centering\epsfig{file=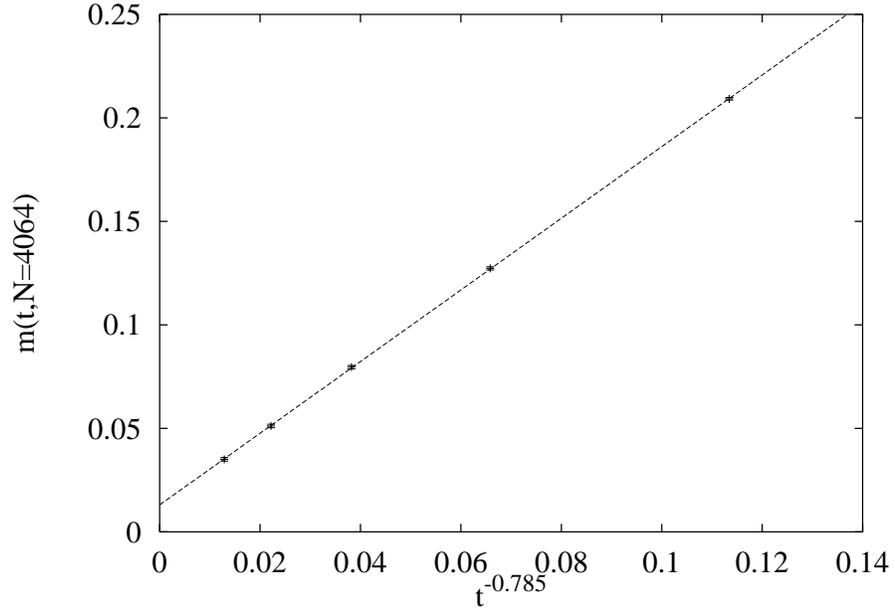,width=0.75\linewidth}
\end{center}
\protect\caption{Remanent magnetization in the SK model at $T=0.8 \,
T_c$. \protect\label{F_m_t}}
\end{figure}

We have repeated this numerical calculation for a temperature below
the critical one ($T=0.8\,T_c$) and sizes $N=480,992,2016,4064$.
In this case we expect that the remanent magnetization tends to a
non-zero asymptotic value due to the finite size of the system.
So we have fitted the data of $M(t,t_w\!=\!3)$ with the following
formula 
\be
m(t,N) = m_\infty(N) + A \, t^{-\lambda(T)}
\ee
where we let $\lambda$ to depend on the temperature.
Via a preliminar three parameter fit we have estimated $\lambda =
0.785(10)$ and with no systematic dependence on the lattice size.
Then fixed the value of $\lambda$ to the just found value, we
extrapolated the value of $m_\infty(N)$ by using a more simple linear
fit, like the one plotted in figure~\ref{F_m_t}.

\begin{figure}
\begin{center}
\leavevmode
\centering\epsfig{file=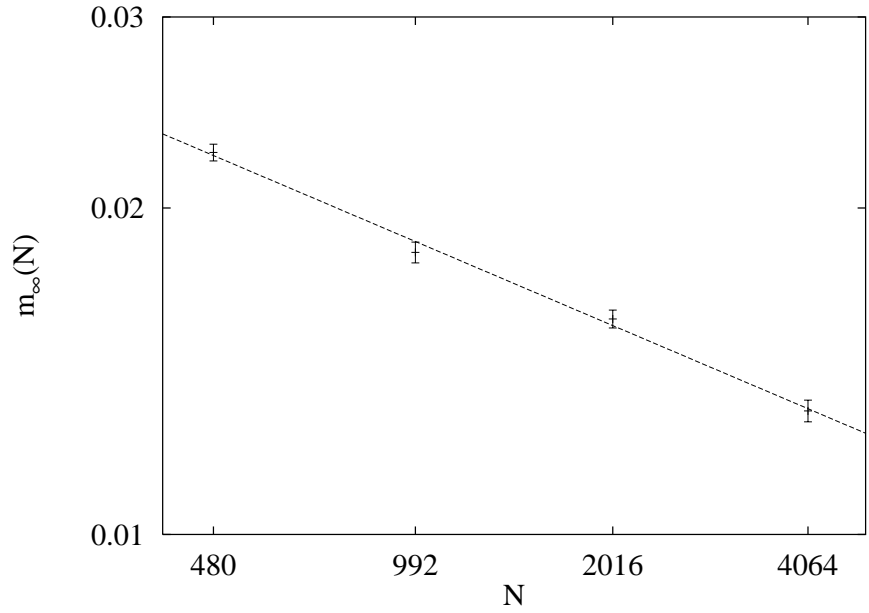,width=0.75\linewidth}
\end{center}
\protect\caption{$m_\infty(N)$ vs. $N$ in the SK model at $T=0.8 \,
T_c$. \protect\label{F_m_inf}}
\end{figure}

Using the values of $m_\infty(N)$ found by the previous analysis we
were able to fit them by a power law of the system size: $m_\infty(N)
\propto N^{-0.25(1)}$. The data with the best fit are reported in a
double-log scale in figure~\ref{F_m_inf} (see ref.~\cite{Rossetti} for
a detailed study).

Assuming a linear dependence of the exponent $\lambda(T)$ with the
temperature, which has been observed in~\cite{Rossetti,PARI} for the
SK model and in~\cite{Johnston} for a spin glass system on quenched
$\phi^3$ graphs (which should behaves like a mean-field SK model), and
from the fact that $\lambda(T=0.8) \simeq 0.8$, we obtain that the
$\lambda$ critical exponent as a function of the temperature is
discontinuous at the critical point (i.e. $\lambda(T_c^-) \simeq 1$
while $\lambda(T_c^+)=5/4$).

\subsection{\protect\label{S_MAGNUM}Numerical estimation in $d=6$ and
$d=8$}

The measurement of the decay rate of the remanent magnetization can be
more difficult than one can imagine, because of the following effect:
we know that $M(t\!=\!0)=1$, but we try to fit the $M(t)$ data with a
power law which diverges at $t=0$.
This effect is evident in a log-log scale where a power fit behaves
like a straight line, while the $M(t)$ data tends to the value 1 when
$\ln t \rightarrow -\infty$; in such a situation we have to discard
the first data points to be sure we are measuring the asymptotic
behavior.
Unfortunately the useless data are just the ones with smaller relative
error, while the data fitted are affected by a greater statistical
indetermination which makes worst the estimation of $\lambda$.

One possible way to overcome this source of error is to measure some
other observable that has the same behavior of $M(t)$, but with a
signal stronger and less fluctuating.
Starting with all the spins up, the magnetization at time $t$ is
nothing more than the overlap between the configuration at time $t$
and the initial one.
If we measure the overlap between the configuration at time $t$ and
the one at a fixed small time ($t_w = 3$ in our case), we expect that
$M(t, t_w)$ behaves like $M(t)$, with similar statistical
fluctuations, but with a signal ten times greater.
 
\begin{figure}
\begin{center}
\leavevmode
\centering\epsfig{file=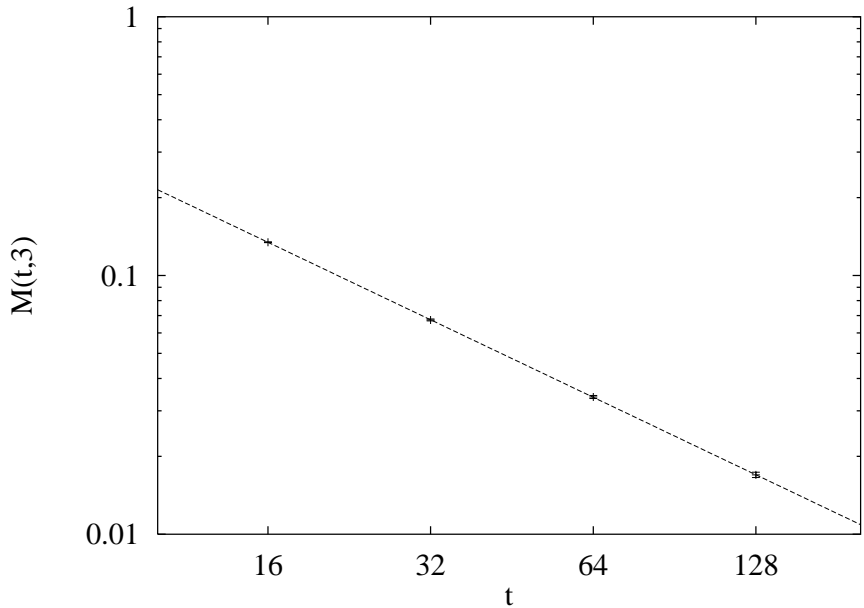,width=0.75\linewidth}
\end{center}
\protect\caption{Remanent magnetization in $d=6$ at $T=T_c$.
\protect\label{F_mag_6d}}
\end{figure}

The results of the simulations of the EA model in 6 dimensions can be
found in figure~\ref{F_mag_6d} where we have plotted $M(t, t_w=3)$ versus
the simulation time $t$; the line is the best power fit which gives an
exponent $\lambda = 0.995 \pm 0.005$.
This value is compatible with 1, but not with the Mean Field value
$\lambda_{\MF} = 5/4$.

As explained in the introduction we believe that the upper critical
dimension becomes $d_u=8$ for such quantities linked, in some sense,
to a magnetic field.
In this case we start the simulation with the system totally
magnetized, like if it was feeling an infinite strength magnetic
field, so the remanent magnetization may be one of such quantities.

Then we have done some simulations of the $\pm1$ Ising Spin Glass
model in $d=8$ to see if we recover the Mean Field behavior of the
remanent magnetization.

The critical temperature in $d=8$ has been extrapolated from the
critical temperatures in $d$=3~\cite{KAWYOU}, 4~\cite{BACIPAPERIRU93}
and 6~\cite{WANG_YOUNG}.
In the limit of $d \rightarrow \infty$ the critical temperature
diverges like $T_c(d) \simeq \sqrt{2d}$.
Moreover, in the Bethe-Peierls approximation, there is an exact
formula for the critical temperature~\cite{MEPA87, PALADIN_SERVA}
\be
(2d-1) \tanh^2(\beta_c) = 1
\label{E_paladin}
\ee
We have used this formula, which turn to be valid in the $d
\rightarrow \infty$ limit, as a starting point adding to it some terms
which may mime the finite dimensions corrections.
In particular we have substitute the r.h.s. of eq.(\ref{E_paladin})
with a fourth order polynomial in $1/d$ (the term of zeroth order
being always 1) and we have tried to fit the known critical
temperatures fixing two term of the polynomial to zero and letting
free the coefficients of the other two terms.
Among all the 6 possible choices we have selected the one with
smaller value of $\chi^2$, which reads
\be
(2d-1) \tanh^2(\beta_c) = 1 + \frac{B}{d^2} + \frac{D}{d^4}
\ee
with $B = 0.95 \pm 0.14$ and $D = 117 \pm 4$.
Through this fit we estimate the critical temperature for the
eight-dimensional EA model as $\beta_c(d=8) = 0.270 \pm 0.001$.
The error reported is an underestimation of the real one because there
are systematic deviations due to the arbitrarily choice of the fitting
function.

\begin{figure}
\begin{center}
\leavevmode
\centering\epsfig{file=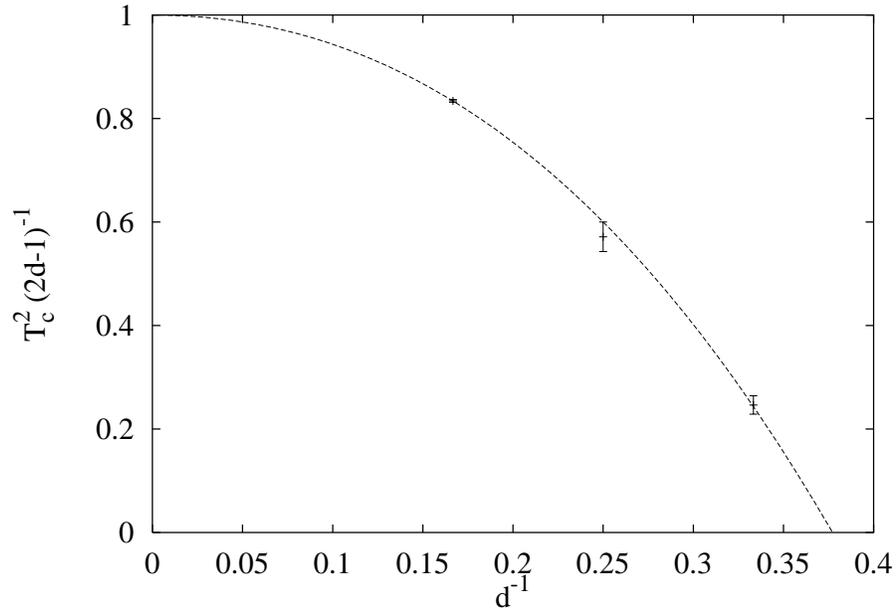,width=0.75\linewidth}
\end{center}
\protect\caption{Polynomial fit to the critical temperatures.
\protect\label{F_beta_c}}
\end{figure}

We have also repeated the analysis looking at the quantity ${T_c}^2 /
(2d-1)$, which takes values in the range $[0,1]$.
Knowing that in the $d \rightarrow \infty$ limit holds that
\be
\frac{{T_c}^2}{2d-1} = 1 \;\;,
\label{E_T_c}
\ee
we have tried to fit the critical temperatures adding to the r.h.s. of
eq.(\ref{E_T_c}) a polynomium in $1/d$, obtaining as the best
resulting interpolation the one reported in figure~\ref{F_beta_c},
which reads 
\be
\frac{{T_c}^2}{2d-1} = 1 - 5.16(26) \frac{1}{d^2} - 5(1) \frac{1}{d^3}
\;\; .
\ee

From the plotted fit we obtain an estimation of $\beta_c(d=8) = 0.270
\pm 0.001$. Also this error should be increased because systematic
deviations due to the arbitrarily choice of the fitting function are
not taken into account.

Anyway the estimations of $\beta_c(d=8)$ obtained with different
interpolations are all in the range $[0.260, 0.270]$ and we think that
the true critical temperature is with very good probability in this
range; in fact, as you can see in figure~\ref{F_beta_c}, the value of
$\beta_c(d=8)$ is strongly dependent on the value of $\beta_c(d=6)$,
which is known with high accuracy, and on the way $\beta_c(d)$ tends
to zero as the dimensionality is increased.

For example if we assume that the successive improvements of the
Bethe-Peierls approximation tend to increase the value of $\beta_c(d)$
for each $d$, then the formula (\ref{E_paladin}) will give a lower
bound for the inverse critical temperature; in $d=8$ this lower bound
is $\beta_c(d=8) > 0.264$.

From figure~\ref{F_beta_c} we can get also another significant
information: the point where the fitting function crosses the x axis
may give us an estimation of the lower critical dimension, which turn
to be $d_l \simeq 2.65$.

\begin{figure}
\begin{center}
\leavevmode
\centering\epsfig{file=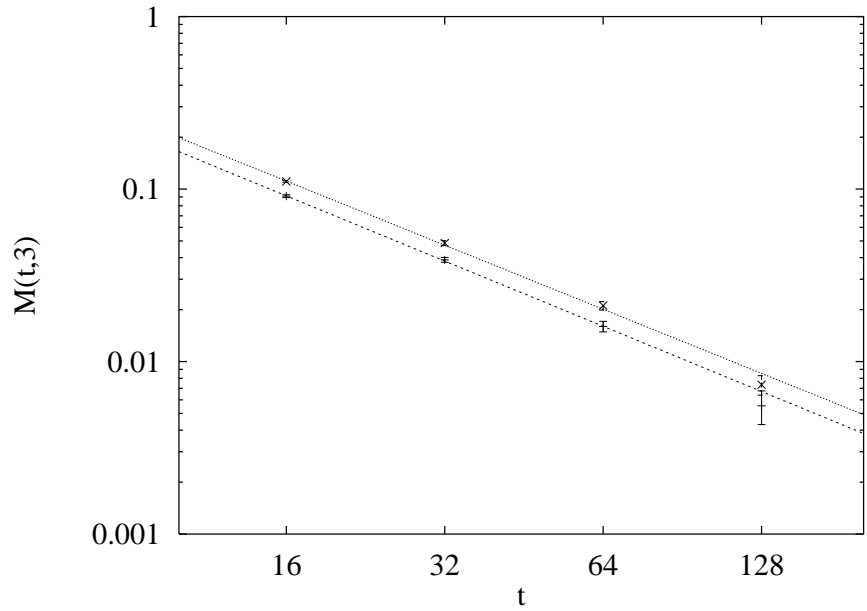,width=0.75\linewidth}
\end{center}
\protect\caption{Remanent magnetization in $d=8$ at $T \simeq T_c$.
\protect\label{F_mag_8d}}
\end{figure}

In figure~\ref{F_mag_8d} we have plotted the data, with the best
power fits, of the remanent magnetization in $d=8$ at temperature
$\beta=0.260$ and $\beta=0.270$.
From the fits we get the exponents $\lambda(\beta\!=\!0.260) = 1.256
\pm 0.08$ and $\lambda(\beta\!=\!0.270) = 1.235 \pm 0.013$, which are
both compatible with the MF result ($\lambda_{\MF} = 5/4$).

\section{\protect\label{S_C}Conclusions}

We have calculated numerically for the first time the dynamical
critical exponents in six dimensions in three different ways, all
compatibles within the statistical error.

Thanks to the previous results we can check also the static critical
exponents (for instance getting the $z$ value obtained from the
scaling of the $q-q$ correlation function), and we obtain values that
agree very well with the static critical exponents and the critical
temperature found in the literature \cite{WANG_YOUNG}.

Finally we have calculated analytically in the Mean Field
approximation the exponent of the remanent magnetization and we obtain
numerical results that confirm that for this observable the upper
critical dimension is eight and not six.

We plain in the future extend this work on the five dimensional Ising
spin glass.

\section{\protect\label{S_ACKNOWLEDGES}Acknowledgments}

We acknowledge interesting discussion with E. Marinari, D. Rossetti
and D. Stariolo. Also, we would like to thank P. Young for provide us
his last estimate on the critical temperature of the four dimensional
$\pm 1$ spin glass.

J .J. Ruiz-Lorenzo is supported by an EC HMC(ERBFMBICT950429) grant.

\end{document}